\documentclass{article}

\usepackage{arxiv}

\usepackage[utf8]{inputenc} 
\usepackage{tabularx} 
\usepackage{pdfpages}
\usepackage{amsmath}  
\usepackage{amssymb} 
\usepackage{amsfonts} 
\usepackage{graphicx} 
\usepackage{wrapfig}
\usepackage{tikz} 
\usetikzlibrary{arrows} 
\usepackage{bm} 
\usepackage{accents} 
\usepackage{pdfpages}
\usepackage{placeins} 
\usepackage{mathtools} 
\usepackage{url}
\usepackage{dsfont}  
\usepackage{algorithm}
\usepackage{algpseudocode}
\usepackage{siunitx}
\usepackage{booktabs,siunitx}
\usepackage{tikz}
\usepackage[round]{natbib}

\title{A Parameter Estimation Method for Multivariate Aggregated Hawkes Processes}

\author{
  Leigh Shlomovich\thanks{Leigh Shlomovich is funded by an Engineering and Physical Sciences Research Council.} \\
  Department of Mathematics\\
  Imperial College London\\
  London, SW7 2AZ\\
  \texttt{leigh.shlomovich14@imperial.ac.uk} \\
   \And
 Edward A. K. Cohen\\
    Department of Mathematics\\
  Imperial College London\\
  London, SW7 2AZ\\
\And
Niall Adams\\
  Department of Mathematics\\
Imperial College London\\
London, SW7 2AZ\\
}

\newcommand{\diff}{{\rm{d}}}

\DeclareMathOperator*{\argmax}{\arg\!\max}
\newcommand{\GG}[1]{}

\begin{document}
\maketitle

\begin{abstract}
It is often assumed that events cannot occur simultaneously when modelling data with point processes. This raises a problem as real-world data often contains synchronous observations due to aggregation or rounding, resulting from limitations on recording capabilities and the expense of storing high volumes of precise data. 
In order to gain a better understanding of the relationships between processes, we consider modelling the aggregated event data using multivariate Hawkes processes, which offer a description of mutually-exciting behaviour and have found wide applications in areas including seismology and finance. 
Here we generalise existing methodology on parameter estimation of univariate aggregated Hawkes processes to the multivariate case using a Monte Carlo Expectation Maximization (MC-EM) algorithm and through a simulation study illustrate that alternative approaches to this problem can be severely biased, with the multivariate MC-EM method outperforming them in terms of MSE in all considered cases.
\end{abstract}


\section{Introduction}

Modern data acquisition systems in many applications collect vast amounts of data that can be characterised as time series. An area of particular interest is cyber-security, where network data are recorded and questions arise about the correlation structure and `excitational' effects in and between the resulting time series. We aim to uncover trends within and between such time stamped event data, which we model using point processes. For this, we are particularly interested in the multivariate Hawkes process, which provides a model for `mutually exciting' events. Introduced in \cite{hawkes_spectra_1971a, hawkes_point_1971b} primarily to model the occurrence of seismic activity, multivariate Hawkes processes have found wide application to many disciplines due to their ability to model cross-excitation. In the case of financial data, multivariate Hawkes processes have been used to model the joint dynamics of trades and mid-price changes of the NYSE \citep{bowsher_modelling_2007} and an overview of literature surrounding the application of these processes to finance is given in \cite{bacry_hawkes_2015}. Additionally Hawkes processes have been considered within cyber-security for modelling of computer network traffic \citep{mark_network_2019, price-williams_nonparametric_2019} and social media activity, for example in \cite{kobayashi_tideh:_2016} where propagation of `Twitter cascades' is considered.

As discussed in \cite{shlomovich_monte_2020} practical limitations on storing or recording high resolution data results in an abundance of aggregated data, that is streams of counts of occurrences per time bin. We can use the counting process representation of a $P$-variate point process to model this aggregated data. Let $N^{(p)}(t)$ be the $p$th continuous time count process ($p= 1, \ldots, P$) denoting the number of events until time $t \in \mathbb{R}$ where $N^{(p)}(0) = 0$, $N^{(p)}(t) = N^{(p)}((0,t])$ for $t > 0$ and $-N^{(p)}((t,0])$ for $t < 0$. We further denote 
\begin{equation*}
N^{(p)}_{t} = N^{(p)}(\Delta(t+1)) - N^{(p)}(\Delta t), 
\end{equation*}
to be the aggregated (binned) process. Then the discrete-time vector process $\bm{N}_t$ is given by 
\begin{equation*}
\bm{N}_t = \left[N^{(1)}_{t}, \ldots, N^{(P)}_{t}\right], \, t = 1, \ldots, K,
\end{equation*}
\sloppy
where $K = T/\Delta$ and $\Delta$ is the bin width. Here we develop and extend the methodology detailed in \cite{shlomovich_monte_2020} to handle parameter estimation of multivariate aggregated Hawkes processes. We do this by considering the superposition of the multivariate count process. This is an elegant method which manages to replicate both the marginal and the cross-correlation structure of the true process. Results from a simulation study, comparing the extended MC-EM algorithm to both the INAR($p$) and binned log-likelihood approximations are presented in Section \ref{sec:sim_study}. This method is shown to produce estimates with lower MSE than currently available alternatives. We additionally derive the Hessian of a multivariate Hawkes process with exponential kernel, required for optimization, and present this in Appendix \ref{app:likelihood} along with the gradient derived in \cite{ozaki_maximum_1979}. 

\subsection{Multivariate Hawkes Processes}\label{sec:mv_hp_theory}

Formally, the $P$-dimensional Hawkes process $\bm{N}(t) = \left[N^{(1)}(t), \ldots, N^{(P)}(t)\right]$ is a class of stochastic process such that independently for each $p \in \{1, \ldots P\}$,
\begin{align*}
\operatorname{Pr}\{\diff N^{(p)}(t)=1 \mid {\bm{N}}(s) \, (s \leq t)\} &= \lambda_{(p)}^{*}(t) \diff t+o(\diff t), \\
\operatorname{Pr}\{\diff N^{(p)}(t)>1 \mid {\bm{N}}(s) \, (s \leq t)\} &= o(\diff t),
\end{align*}
where $\mathrm{d} N^{(p)}(t)=N^{(p)}(t+\diff t)-N^{(p)}(t)$ \citep{hawkes_spectra_1971a}. It is characterized via its conditional intensity function (CIF) $\lambda_{(p)}^{*}(t),$ defined as
\begin{equation}\label{eqn:cif1}
\lambda_{(p)}^{*}(t)=\nu_p + \sum_{m=1}^{P} \int_{-\infty}^{t} g_{pm}(t-u) \diff N^{(m)}(u),
\end{equation}
where $\bm{\nu}>0$ is a $P$-dimensional vector called the background intensity and $\bm{g}(u)$ is the non-negative excitation kernel such that  $\bm{g}(u) = 0$ for $u < 0$ and given by a $P \times P$ matrix of functions. In this way, the intensity at an arbitrary time-point is dependent on the history of the multivariate process allowing for both self and mutually exciting behavior. In simple cases, if $g_{ij}(u)=0$ for all $u$ and $i \neq j$, then this is non-cross exciting behavior. That is the cross-covariances are equal to zero and whilst the processes may be self exciting, they are independent and thus not mutually exciting \citep{hawkes_spectra_1971a}. Further, in the bivariate case if $g_{21}(u)=0$ for all $u$ but $g_{12} (u) \neq 0$ for all $u$ then $N^{(1)}(t)$ does not affect the likelihood of events in $N^{(2)}(t)$, but the converse does not hold. In this case we have one way interaction between the two processes involved. 

The CIF from (\ref{eqn:cif1}) can be written for an exponential kernel as
\begin{equation*}
\lambda_{(p)}^{*}(t)=\nu_{p}+\sum_{m=1}^{P} \sum_{j=1}^{N^{m}(t)} \alpha_{ p m} \exp \left(-\beta_{p m}\left(t-t^{m}_{j}\right)\right),
\end{equation*}
where $\bm{\nu}$, $\bm{\alpha} = \left( \alpha_{pm} \right)$, $\bm{\beta} = \left( \beta_{pm} \right)$, $p,m = 1, \ldots P$ are known as the baseline, excitation and decay parameters, respectively. Assuming stationarity of the multivariate Hawkes process we have that the vector of stationary densities is 
\begin{equation*}
\bm{\lambda} = E \{ \bm{ \lambda}^* \}
\end{equation*}
so that
\begin{align*}
\bm{\lambda} &=\left(\bm{I}_{P}-\int_{0}^{\infty} \bm{g}(u) \diff u\right)^{-1} \bm{\nu} =\left(\bm{I}_{P}-\bm{G}(0)\right)^{-1} \bm{\nu},
\end{align*}
where $\boldsymbol{G}(\omega)$ is the Fourier transform of the excitation kernel $\bm{g}(\cdot),$ given by
\begin{equation*}
\bm{G}(\omega)=\int_{-\infty}^{\infty} e^{-i \omega \tau} \bm{g}(\tau) \mathrm{d} \tau.
\end{equation*}
Note that $\bm{G}(0)$ is commonly referred to as the branching ratio, typically denoted by 
$\bm{\gamma}$, with the condition for stationarity being that the spectral radius $\rho(\bm{\gamma})<1$ \citep{hawkes_point_1971b}.

\section{Multivariate MCEM for Aggregated Hawkes Processes}

In a continuous time framework, maximum likelihood estimation (MLE) can be used to estimate the model parameters from a set of exact multivariate events on the interval $[0,T]$ \citep{ozaki_maximum_1979} denoted 
\begin{equation*}
\bm{\mathcal{T}} = \Big\{ \mathcal{T}_{(p)} \Big\}_{p=1, \ldots P} = \Big\{ t^p_1, \ldots t^p_{N^{(p)}_T} \Big\}_{p=1, \ldots P} \in [0, T],
\end{equation*}
where $T$ is the maximum observation or simulation time, $t^p_l$ is the $l^{\rm{th}}$ event in process $p$, and $N^{(p)}(T)$ is the total number of events in process $p$. When we observe an aggregation of these latent continuous times to a count process of events per time bin, we lose information and in particular the likelihood of our observed event times given a parameter set can no longer be computed exactly due to reliance on the underlying time-stamps. In particular, the Hawkes process which is defined by its CIF, depends on the history of the process, which if `blurred' by binning or rounding requires correct handling in order to obtain meaningful parameter estimates. In \cite{shlomovich_monte_2020} an MC-EM (Monte Carlo Expectation Maximisation) algorithm for the parameter estimation of univariate aggregated Hawkes processes is presented. Here this work is extended for multivariate data.


\subsection{The Monte Carlo EM Algorithm }\label{sec:MCEM}

\sloppy
The EM algorithm \citep{dempster_maximum_1977} augments observed data by a latent quantity \citep{wei_monte_1990} to iteratively compute the maximizer of a likelihood. Here, the observed data are the multivariate event counts per unit time 
$
\bm{N}_t = [N^{(1)}_{t}, \ldots, N^{(P)}_{t}].
$
The latent data, $\bm{\mathcal{T}}$ are the unobserved, true event times. These are marked time-stamps which are not observed due to practical restrictions. We denote the parameter set to be $\bm{\Theta}=\{\bm{\nu}, \bm{\alpha}, \bm{\beta}\}$, where $\bm{\nu}$ is $P \times 1$, and $\bm{\alpha}$ and $\bm{\beta}$ are $P \times P$. The following two steps are as detailed in \cite{shlomovich_monte_2020}: 
\begin{enumerate}
	\item In the E (Expectation) step, we compute
	\begin{align}
	\label{eqn:exactQ} 
	Q_{i+1}(&\bm{\Theta}, \bm{\Theta}^i) = \nonumber \\
	&\int_{\mathbb{T}} \log (p (\bm{\Theta} \mid {\bm{N}}, \bm{\mathcal{T}})) p ( \bm{\mathcal{T}} \mid {\bm{N}}, \bm{\Theta}^i) \diff \bm{\mathcal{T}},
	\end{align}
	where $\mathbb{T}$ denotes the sample space for $\bm{\mathcal{T}}$. 
	\item In the M (Maximization) step, maximize the conditional expectation in (\ref{eqn:exactQ}) to obtain the updated parameter estimate, $\bm{\Theta}^{i+1}$.
\end{enumerate}
Monte Carlo methods can be used to numerically compute (\ref{eqn:exactQ}) if it is intractable, forming an algorithm known as MCEM \citep{wei_monte_1990}. As we cannot sample the time-stamps directly we use importance sampling to simulate proposals $\bm{\mathcal{T}}^*$ for $\bm{\mathcal{T}}$ from a feasible alternative distribution, denoted $q(\bm{\mathcal{T}} \mid {\bm{N}}, \bm{\Theta}^i)$. 

Unique to the multivariate formulation is the need to retain the latent covariance structure between the $P$ processes. To this end, we sample the times of the univariate superposition of the $P$-variate process. We then split the superposed simulated time-stamps to $P$ processes to create proposals matching the multivariate counts. We refer to such viable proposals as {\it{consistent}}. This approach is detailed further in Section \ref{sec:superpose}. \sloppy Once sampled, each proposal is weighted according to the probability it came from the desired distribution. That is, given a set of $M$ samples $\bm{\mathcal{T}}^{\ast (1)},\ldots,\bm{\mathcal{T}}^{\ast (M)}$, we assign weights
\begin{align}\label{eqn:weights}
w_k = \frac{p(\bm{\mathcal{T}}^{\ast (k)} \mid {\bm{N}}, \bm{\Theta}^i)}{q(\bm{\mathcal{T}}^{\ast (k)} \mid {\bm{N}}, \bm{\Theta}^i)},
\end{align}
and approximate (\ref{eqn:exactQ}) with
\begin{align}
\label{eqn:impQ}
Q_{i+1}(\bm{\Theta}, \bm{\Theta}^{i}) = \frac{\sum_{k=1}^M w_k \log (p (\bm{\Theta} \mid {\bm{N}}, \bm{\mathcal{T}}^{\ast (k)}))}{ \sum_{k=1}^M w_k  }.
\end{align}
It is shown in \cite{shlomovich_monte_2020} that if only proposing consistent event times
\begin{equation*}
p \left( \bm{\mathcal{T}}^{\ast (k)} \mid {\bm{N}, \bm{\Theta}^i} \right) \propto p \left( \bm{\mathcal{T}}^{\ast (k)} \mid \bm{\Theta}^i \right),
\end{equation*}
where $\log \left( p(\bm{\mathcal{T}}^{\ast (k)}\mid \bm{\Theta}) \right)$ is given in \cite{daley_introduction_2003} by
\begin{align*}
\log \mathcal{L}&(\bm{\Theta} ; \bm{\mathcal{T}}) = \sum_{p=1}^{P} \log \mathcal{L}^p(\bm{\Theta} ; \bm{\mathcal{T}}), \\
&= \sum_{p=1}^P \left[ \sum_{j=1}^{N^{(p)}(T)} \log \lambda_{(p)}^{*}\left(t^{(p)}_{j}\right)-\int_{0}^{T} \lambda_{(p)}^{*}(u) \mathrm{d} u. \right].
\end{align*}


\subsection{Multivariate Sampling via the Superposition Process}\label{sec:superpose}
\sloppy
Generating consistent time-stamps for the latent multivariate process such that the true covariance structure is captured is an important problem. We propose the following method for this, which allows us to extend the work in \cite{shlomovich_monte_2020}
for multivariate count data. Consider $\bm{N}_t=[N_t^{(1)}, \ldots, N_t^{(P)}]$ being a binned $P$-variate Hawkes process with exponential kernel. In order to generate possible multivariate, continuous time proposal, $\bm{\mathcal{T}}^{\ast}$, of the underlying event times, $\bm{\mathcal{T}}$, we consider the superposed binned count process, 
\[
\tilde{N}_t=N^{(1)}_t+ \ldots + N^{(P)}_t,
\]
and reparameterize the multivariate parameter set $\bm{\Theta}=\{\bm{\nu}, \bm{\alpha}, \bm{\beta}\}$ to a corresponding set for the univariate superposed process $\tilde{N}$, according to the kernel. We denote the adjusted parameter set for $\tilde{N}$ as $\tilde{\Theta} = \{\tilde{\nu}, \tilde{\alpha}, \tilde{\beta}\} $.

It is given that the intensity of the superposed count process $\tilde{N}$ can be written as $\tilde{\lambda}(t) = \sum_{p=1}^{P} \lambda_{(p)}^*(t)$. The stationary intensity of the consistent proposed time-stamps is equal to that of the superposed observed counts, and so we have that
\begin{equation}\label{eqn:tilde_relation}
\frac{\tilde{\nu}}{ 1- \tilde{\gamma}} = \frac{1}{K \Delta} \sum_{p=1}^P E\{ N^{(p)} (T)\},
\end{equation}
where $K = T/\Delta$ and $\tilde{\gamma}$ is the branching ratio of the superposed process, equal to $\tilde{\alpha}/\tilde{\beta}$ in the case of an exponential kernel. We additionally note that $\tilde{\nu} = \sum_{p=1}^P \nu_p$. This can be seen by noting that 
\begin{align*}
\tilde{\lambda}(t) &= \sum_{p=1}^{P} \lambda_{(p)}^*(t),\\
\implies \tilde{\lambda}(t) &= \sum_{p=1}^{P} \nu_p \\ 
&\; \quad +  \sum_{p=1}^{P}  \sum_{m=1}^{P} \sum_{j=1}^{N^{m}(t)} \alpha_{pm} \exp \left(-\beta_{pm}\left(t-t^{m}_{j}\right)\right),\\
&\equiv \tilde{\nu} + \sum_{j=1}^{\tilde{N}(t)} \tilde{\alpha} \exp \left(-\tilde{\beta} \left(t-t_{j}\right)\right).
\end{align*}
It remains to define a reparameterization for $\tilde{\alpha}$ and $\tilde{\beta}$. From Equation (\ref{eqn:tilde_relation}) and noting that $\tilde{\nu} = \sum_{p=1}^P \nu_p$, we have 
\begin{equation}\label{eqn:gamma_reparam}
\tilde{\gamma} = 1 - \frac{K \Delta  \sum_{p=1}^P \nu_p}{ \sum_{p=1}^P E\{ N^{(p)} (T)\} }.
\end{equation}
Here we consider the exponential kernel where $\tilde{\gamma} = \tilde{\alpha}/\tilde{\beta}$ and thus we can use Equation (\ref{eqn:gamma_reparam}) in order to retain a consistent stationary intensity in the univariate proposal.  

We find that a suitable choice for $\tilde{\beta}$ is
$$
\tilde{\beta}=\operatorname{mean}\left(\beta_{i j}\right)_{i, j},
$$
for $i,j = 1, \ldots P$, and by Equation (\ref{eqn:gamma_reparam})
$$
\tilde{\alpha}= \operatorname{mean}\left(\beta_{i j}\right)_{i, j} \left( 1 - \frac{K \Delta \sum_{p=1}^P \nu_p}{ \sum_{p=1}^P E\{ N^{(p)} (T) \} } \right).
$$
Empiricial studies show this reparameterization accurately recovers the true CIF of the superposed process. In this way we only simulate a univariate process, albeit a superposed version, and denote the simulated times as $\tilde{\mathcal{T}}^{\ast}$. In order to generate a realization of the multivariate Hawkes process, $\bm{\mathcal{T}}^{\ast}$, with cross-covariances, we uniformly sample the observed number of events in each bin for each process from the $\tilde{\mathcal{T}}^{\ast}$. That is, we simulate a consistent set of continuous times for the superposed process matching the observed counts for $\tilde{N}$, and then uniformly assign points within each bin to each of the $P$ processes. If desired, the allocation of events to the $P$ processes can be conducted $\tilde{m}$ times to generate multiple possible multivariate versions of the proposed realization of $\tilde{N}$. If $\tilde{m}>1$, the sample which maximises the log-likelihood is selected, otherwise the single multivariate proposal is taken and the MLE is used to estimate the parameters of the multivariate process from the continuous-time proposed realization. We present results for $\tilde{m}=10$ in Section \ref{sec:sim_study}.

This method provides us with an efficient way of artificially injecting cross-correlation into the consistent proposals whilst also retaining the marginal properties. In order to speed up the maximisation of the likelihood, we require the gradient and Hessian, given in Appendix \ref{app:likelihood}. The full algorithm for this multivariate approach is derived and presented in Appendix \ref{app:mcem_alg}.

\subsection{Sampling Method}\label{sec:sample_method}
\sloppy
The question remains of how to best sample the latent times. The sequential simulation method detailed in \cite{shlomovich_monte_2020} is applicable in the multivariate extension due to the reparameterization step meaning we need only sample times for the univariate superposition process. In this case, the normalised weight of the $k$th Monte Carlo sample ($k=1, \ldots, M$) is given by
\begin{equation}\label{eqn:og_weights}
w_{k} = \frac{\splitfrac{ \exp \left(\log \left( p\left(\bm{\mathcal{T}}^{*(k)} \mid \bm{N}, \bm{\Theta}^{i}\right) \right) \right.} { \left. - \log \left( q\left(\bm{\mathcal{T}}^{*(k)} \mid \bm{N}, \bm{\Theta}^{i} \right) \right)\right) } } 
{ \splitfrac{ \sum_{k=1}^{M}\left( \exp \left( \log \left( p \left(\bm{\mathcal{T}}^{*(k)} \mid \bm{N}, \bm{\Theta}^{i}\right) \right) \right) \right.}{ \left. - \log \left( q\left(\bm{\mathcal{T}}^{*(k)} \mid \bm{N}, \bm{\Theta}^{i}\right)\right) \right) } },
\end{equation}
where $M$ is the number of Monte Carlo samples. Further we note that
\begin{align*}
\log \left( q\left(\bm{\mathcal{T}}^{*(k)} \mid \bm{N}, \bm{\Theta}^{i}\right)\right) = 
\log \left( q\left(\tilde{\bm{\mathcal{T}}}^{*(k)} \mid \tilde{N}, \tilde{\Theta}^{i}\right)\right) \\
\quad + \log \left( {\rm{Pr}} \left( \mathcal{T}^{*(k)} \mid  \tilde{\bm{\mathcal{T}}}^{*(k)} , {\bm{N}} \right) \right),
\end{align*}
where 
$
\log \left( q\left(\tilde{\bm{\mathcal{T}}}^{*(k)} \mid \tilde{N}, \tilde{\Theta}^{i}\right)\right)
$
is the log-likelihood of the sequentially sampled superposed times given the superposed counts and reparameterized univariate estimates, and
$
{\rm{Pr}} \left( \mathcal{T}^{*(k)} \mid  \tilde{\bm{\mathcal{T}}}^{*(k)} , {\bm{N}} \right)
$
denotes the probability of the random division of superposed time-points into a $P$-variate count process matching the observed counts. 

Large differences between $p\left(\bm{\mathcal{T}}^{*(k)} \mid \bm{N}\right)$ and $q\left(\bm{\mathcal{T}}^{*(k)} \mid \bm{N}\right)$ can result in the weights being close to zero. Therefore we rescale the exponent term to avoid arithmetic underflow when computing the weights. We have
\begin{align}\label{eqn:rescaled_weights}
&\frac{\splitfrac{ \exp \left(\log \left( p\left(\bm{\mathcal{T}}^{*(k)} \mid \bm{N}, \bm{\Theta}^{i}\right) \right) \right.} { \left. - \log \left( q\left(\bm{\mathcal{T}}^{*(k)} \mid \bm{N}, \bm{\Theta}^{i} \right) \right) - C\right) } } 
{ \splitfrac{ \sum_{k=1}^{M}\left( \exp \left( \log \left( p \left(\bm{\mathcal{T}}^{*(k)} \mid \bm{N}, \bm{\Theta}^{i}\right) \right) \right) \right.}{ \left. - \log \left( q\left(\bm{\mathcal{T}}^{*(k)} \mid \bm{N}, \bm{\Theta}^{i}\right)\right) - C \right) } },\\
&=\frac{\splitfrac{ \exp \left(\log \left( p\left(\bm{\mathcal{T}}^{*(k)} \mid \bm{N}, \bm{\Theta}^{i}\right) \right) \right.} { \left. - \log \left( q\left(\bm{\mathcal{T}}^{*(k)} \mid \bm{N}, \bm{\Theta}^{i} \right) \right)\right)/ \exp (C) } } 
{ \splitfrac{ \sum_{k=1}^{M}\left( \exp \left( \log \left( p \left(\bm{\mathcal{T}}^{*(k)} \mid \bm{N}, \bm{\Theta}^{i}\right) \right) \right) \right.}{ \left. - \log \left( q\left(\bm{\mathcal{T}}^{*(k)} \mid \bm{N}, \bm{\Theta}^{i}\right)\right) / \exp (C) \right) } }, \nonumber\\
&=w_{k} \nonumber .
\end{align}
Thus we can use Equation (\ref{eqn:rescaled_weights}) in place of Equation (\ref{eqn:og_weights}). 

\section{Simulation Study} 
\label{sec:sim_study}
\sloppy
We conduct a simulation study to compare the performance of the multivariate MC-EM algorithm to the INAR($p$) method introduced in \cite{kirchner_hawkes_2016}. We also compare the results to an approximation which ignores inter-bin excitation, referred to as the binned log-likelihood method. This approach represents the CIF as a piecewise constant function within each bin, equivalently assuming that $N_j^{(p)} \sim {\rm{Poisson}}  \{ \Delta \lambda^*_{m} \left( [j-1] \Delta \right) \}$, where $N_j^{(p)}$ are the counts in the $j$th bin of the $p$th process \citep{mark_network_2019}. 

Given parameters $\bm{\nu},$ a $2 \times 1$ matrix, and $\bm{\alpha}$ and $\bm{\beta}$ both being $2 \times 2$ matrices, along with some maximum simulation time $T$, we can simulate realizations of a Hawkes process. The generated events $\bm{\mathcal{T}}$ represent the underlying process and aggregating these to a chosen binning $\Delta$ allows us to simulate the count data $\{\bm{N}_t, \, t=1, \ldots, K  \}$. We then apply each of the multivariate MC-EM methods, INAR($p$) and binned log-likelihood approximation. 

\begin{figure*}[htb]
	\hspace{0.5cm} \makebox[\textwidth][c]{\includegraphics[scale=1]{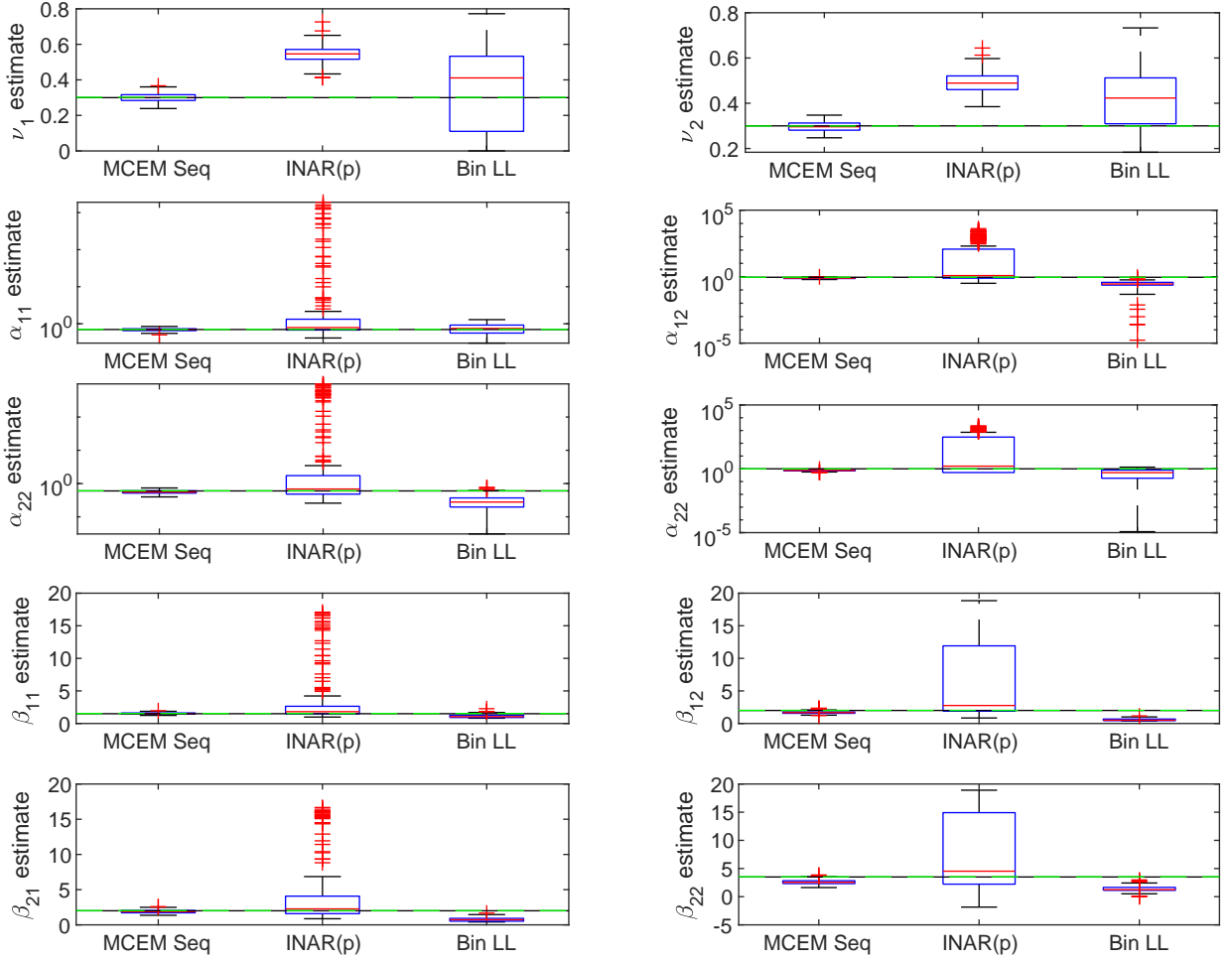}}%
	\caption{For comparison, the green dashed lines represent the mean MLE of the ground truth continuous times and the black solid lines the ground truth itself. Note, the INAR($p$) method can also produce negative values which are not shown in the case of log scales being used. Log scales have been used in any figure where a parameter estimate was greater than or equal to 1000.}
	\label{fig:mv_boxplot}
\end{figure*}

Boxplots for each of the ten estimated parameters used for characterising the bivariate Hawkes process are given in Figure \ref{fig:mv_boxplot}. The parameters used for simulation are
\begin{equation*}
\bm{\nu}=
\left[ {\begin{array}{c}
	0.3 \\
	0.3 \\
	\end{array} } \right], \;  
\bm{\alpha}=
\left[ {\begin{array}{cc}
	0.7 & 0.9 \\
	0.6 & 1.0 \\
	\end{array} } \right], \; 
\bm{\beta}=
\left[ {\begin{array}{cc}
	1.5 & 2.0 \\
	2.0 & 3.5 \\
	\end{array} } \right], 
\end{equation*}
with $ \Delta = 1$ and $T = 2000$. The parameters have been chosen as a stationary case with ample cross-excitation and non-symmetric self-excitation. The mean parameter estimates from repeated simulations is presented on the vertical axis. The INAR($p$) approximation method can yield highly variable results, which is to be expected as the method is primarily intended for selecting a parametric kernel from continuous time data. The usual implementation of the INAR($p$) involves selecting $\Delta$ such that there is approximately one event per bin, however for our application this is not possible and so the choice of $\Delta$ here is chosen to better reflect real data. A log-scale has been used in all four graphs relating to each of $\bm{\alpha}$ and $\bm{\beta}$. Both the binned log-likelihood, and particularly the INAR($p$) method produced large outliers, resulting in very large MSE relative to the MC-EM approach. Tables \ref{table:rel_bias} and \ref{table:trim_mean} present summary statistics for the outlier-trimmed data. Specifically we remove the top and bottom 5\% of parameter estimates for each of the three methods and report the relative bias, mean and standard deviation for each of the parameters and methods. We see that even after removing outliers and negative values from the INAR$(p)$ parameter estimates, the variability is much higher than that of the MC-EM. The absolute value of the bias of both the INAR$(p)$ and binned log-likelihood approaches is also larger than that of the MC-EM, in some cases significantly so. Overall, the proposed MCEM approach has much improved estimation performance than the other methods. 

\begin{table*}
	\centering
	\addtolength{\tabcolsep}{-0.3pt}
	\begin{tabular}{
			@{}
			l
			S[table-format=1.2]
			S[table-format=1.2]
			S[table-format=1.2]
			@{}
		}
		\toprule
		{Parameter} & {MCEM Rel. Bias} & {INAR($p$) Rel. Bias} & {Binned LL Rel. Bias}\\
		\midrule
		$\nu(1)$ & -0.003 &  0.813 & 0.117 \\ 
		\hline
		$\nu(2)$ & -0.010 & 0.633 & 0.363 \\ 
		\hline
		$\alpha(1,1)$ & -0.007 & 22.427 & 0.066 \\ 
		\hline		
		$\alpha(1,2)$ & -0.140 & 204.556 & -0.674\\ 
		\hline
		$\alpha(2,1)$ & -0.0683 &  64.167 & -0.520 \\ 
		\hline
		$\alpha(2,2)$ & -0.257 & 142.000 & -0.510 \\ 
		\hline
		$\beta(1,1)$ & 0.013 & 0.847 & -0.273\\ 
		\hline
		$\beta(1,2)$ & -0.160 &  2.010 & -0.725\\ 
		\hline
		$\beta(2,1)$ & -0.0600 & 0.915 & -0.632\\ 
		\hline
		$\beta(2,2)$ & -0.2660 &  1.217 & -0.609 \\
		\bottomrule
	\end{tabular}
	\caption{Relative bias for the data trimmed, removing the top and bottom 5\% of values for each of the three considered methods.}
	\label{table:rel_bias}
\end{table*}

\begin{table*}[ht!]
	\setlength{\tabcolsep}{4pt}
	\centering
	\begin{tabular}{@{}lSSSSSSSSS@{}}
		\toprule
		& \multicolumn{1}{c}{Ground Truth} & \multicolumn{2}{c}{\hspace{-0.15cm} MCEM Mean (sd)} & \multicolumn{2}{c@{}}{\hspace{-0.15cm} INAR($p$) Mean (sd)} & \multicolumn{2}{c@{}}{\hspace{-0.15cm} Binned LL Mean (sd)}  & \multicolumn{2}{c@{}}{\hspace{-0.15cm} MLE Mean (sd)} \\
		\midrule
		$\nu_{1}$ & 0.30 & 0.30 & \hspace{-0.3cm} (0.02) & 0.54 & \hspace{0cm} (0.04) & \hspace{0.6cm} 0.34 & \hspace{-0.1cm} (0.20) & 0.30 & \hspace{-0.3cm} (0.02)\\
		\hline
		$\nu_{2}$ & 0.30 & 0.30 & \hspace{-0.3cm} (0.018) & 0.490 & \hspace{0cm} (0.036) & \hspace{0.6cm} 0.409 & \hspace{-0.1cm} (0.099) & 0.299 & \hspace{-0.3cm} (0.016)\\
		\hline
		$\alpha(1,1)$ & 0.70 & 0.695 & \hspace{-0.3cm} (0.05) & 16.40 & \hspace{-0.15cm}  (79.40) & \hspace{0.6cm} 0.75 & \hspace{-0.1cm} (0.20) & 0.71 & \hspace{-0.3cm} (0.05)\\
		\hline
		$\alpha(1,2)$ & 0.90 & 0.77 & \hspace{-0.3cm} (0.06) & 185.00 & \hspace{-0.3cm} (394.00) & \hspace{0.6cm} 0.29 & \hspace{-0.1cm} (0.09) & 0.91 & \hspace{-0.3cm} (0.08)\\
		\hline
		$\alpha(2,1)$ & 0.60 & 0.56 & \hspace{-0.3cm} (0.05) & 39.10 & \hspace{-0.3cm} (127.00) & \hspace{0.6cm} 0.29 & \hspace{-0.1cm} (0.11) & 0.61 & \hspace{-0.3cm} (0.06)\\
		\hline
		$\alpha(2,2)$ & 1.00 & 0.74 & \hspace{-0.3cm} (0.06) & 143.00 & \hspace{-0.3cm} (236.00) & \hspace{0.6cm} 0.49 & \hspace{-0.1cm} (0.33) & 0.99 & \hspace{-0.3cm} (0.09)\\		
		\hline
		$\beta(1,1)$ & 1.50 & 1.52 & \hspace{-0.3cm} (0.12) & 2.77 & \hspace{0cm} (2.72) & \hspace{0.6cm} 1.09 & \hspace{-0.1cm} (0.18) & 1.53 & \hspace{-0.3cm} (0.11)\\
		\hline
		$\beta(1,2)$ & 2.00 & 1.68 & \hspace{-0.3cm} (0.14) & 6.02 & \hspace{0cm} (5.56) &\hspace{0.6cm}  0.55 & \hspace{-0.1cm}  (0.14) & 2.01 & \hspace{-0.3cm} (0.18)\\
		\hline
		$\beta(2,1)$ & 2.00 & 1.88 & \hspace{-0.3cm} (0.19) & 3.83 & \hspace{0cm} (3.90) & \hspace{0.6cm} 0.73 & \hspace{-0.1cm}  (0.21) & 2.01 & \hspace{-0.3cm} (0.19)\\
		\hline
		$\beta(2,2)$ & 3.50 & 2.57 & \hspace{-0.3cm} (0.29) & 7.78 & \hspace{0cm} (5.93) & \hspace{0.6cm} 1.37 & \hspace{-0.1cm}  (0.32) & 3.53 & \hspace{-0.3cm} (0.41)\\
		\bottomrule
	\end{tabular}
	\caption{Mean and standard deviation values for the data trimmed to remove the top and bottom 5\% of values each, to handle outliers. }
	\label{table:trim_mean}
\end{table*}

Evaluation of the methods is additionally explored via goodness of fit. This is an important aspect which allows us to check the validity of the estimates given the data. Often, the random change theorem, given in \cite{daley_introduction_2003}, is used for considering goodness of fit by transforming time-points using the compensator function 
\[
\Lambda(t^p_k) = \int_{0}^{t^p_k} \lambda_{(p)}^{*}(u) \diff u,
\]
where ${t^p_k}$ is the $k^{\rm{th}}$ event in process $p$. In practice, $\lambda_{(p)}^{*}(u)$ is estimated using parameter estimates $\hat{\bm{\Theta}}$, and goodness of fit is conducted by considering the distribution of the transformed times, defined as $\mathcal{T}^{\dagger} = \{t^{\dagger}_1, t^{\dagger}_2, \ldots \} = \{\Lambda(t_1), \Lambda(t_2), \ldots \}$. By the random time change theorem, $\mathcal{T}^{\dagger}$ is a realization of a unit rate Poisson process if and only if $\mathcal{T}$ is a realization from the point process defined by $\Lambda(\cdot)$.

\begin{figure*}[htb]
	\begin{center}
		\includegraphics[scale=0.45]{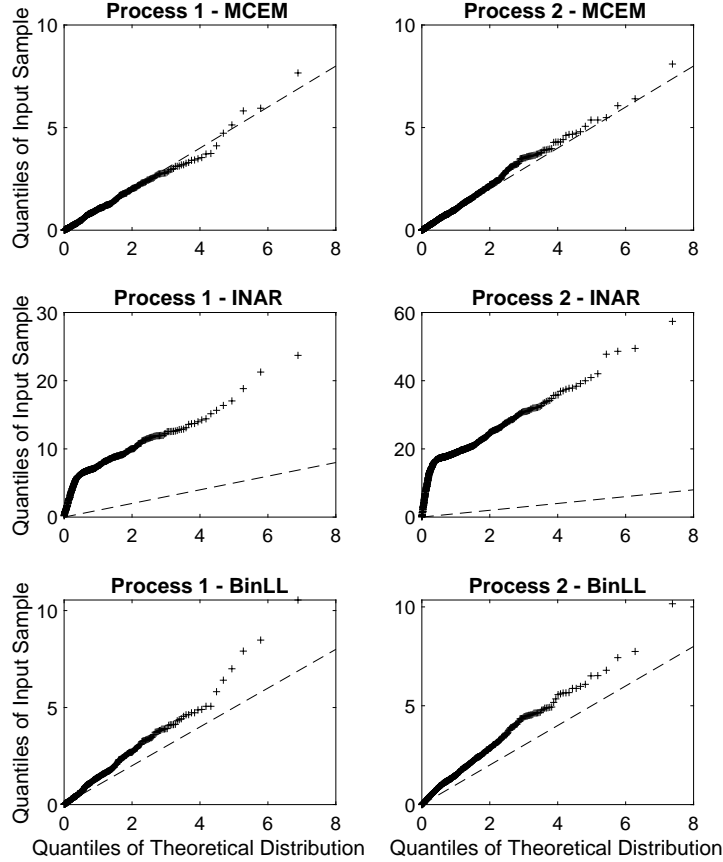}
		\caption{Goodness of fit plots show the MCEM method to yield the most viable parameters.}
		\label{fig:gof}
	\end{center}
\end{figure*}

Figure \ref{fig:gof} shows QQ-plots relating to count processes from the simulation study and comparing the distribution of time-points transformed by each of the parameter estimates against those using the simulation parameters. We see that the estimates produced by the MC-EM algorithm are indeed a viable parameter set for this realization under an exponential Hawkes model, with the transformed time-points being distributed very close to those of the latent times. We also note that as the binned log-likelihood method ignores the effect of inter-bin excitation, it is expected that as the average number of counts in a bin increases, the fit from using this method will worsen.

\section{Case Study}

%
%

Network flow data, referred to as NetFlow, assembles records exported by routers and describes communications between devices connected to an enterprise network. Monitoring and analysing NetFlow data has been successful at detecting a range of malicious network behavior \citep{turcotte_unified_2018}. Here we detect both self-exciting effects in the communication between a pair of network devices, termed here as an 'edge', and mutually-exciting activity between such edges in the Los Alamos National Lab (LANL) enterprise network. By modelling the activity in this way, insight into the correlation structure of communications in the network can be obtained and monitored. We consider the methods outlined in Section \ref{sec:sim_study} for parameter estimation of aggregated Hawkes processes as the NetFlow data is recorded at a 1 second resolution resulting in multiple events occurring simultaneously. 
\begin{figure*}[!htb]
	\centering
	\includegraphics[scale=0.4]{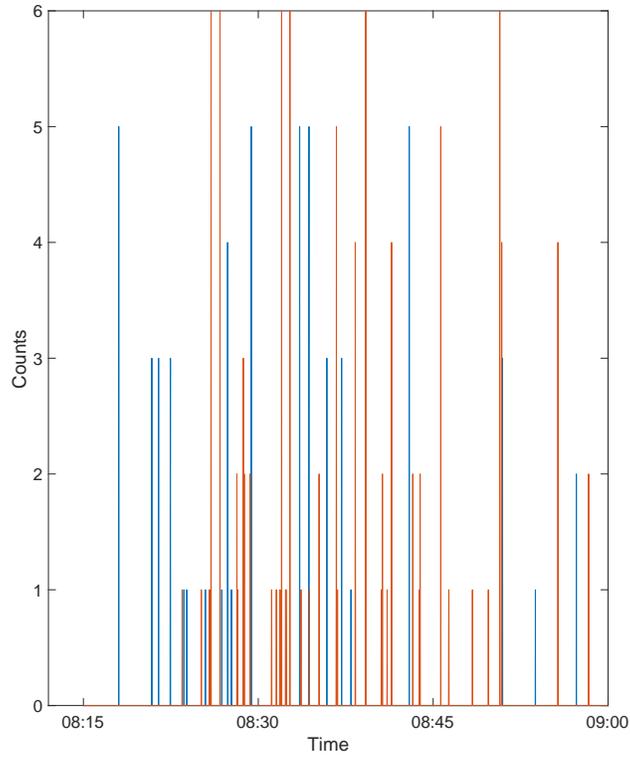}
	\caption{Counts of NetFlow event data on 2 edges in the LANL network. }
	\label{fig:10}
\end{figure*}
Figure \ref{fig:10} presents a selected pair of edges in the LANL network with possibe mutually-exciting Hawkes behavior in the communications over a duration of 45 minutes. We refer to the counts in blue as process 1 and the counts in red as process 2. The window selected is chosen as a period of more frequent events surrounded by no activity. 
\begin{figure*}[!htb]
	\centering
	\includegraphics[scale=0.4]{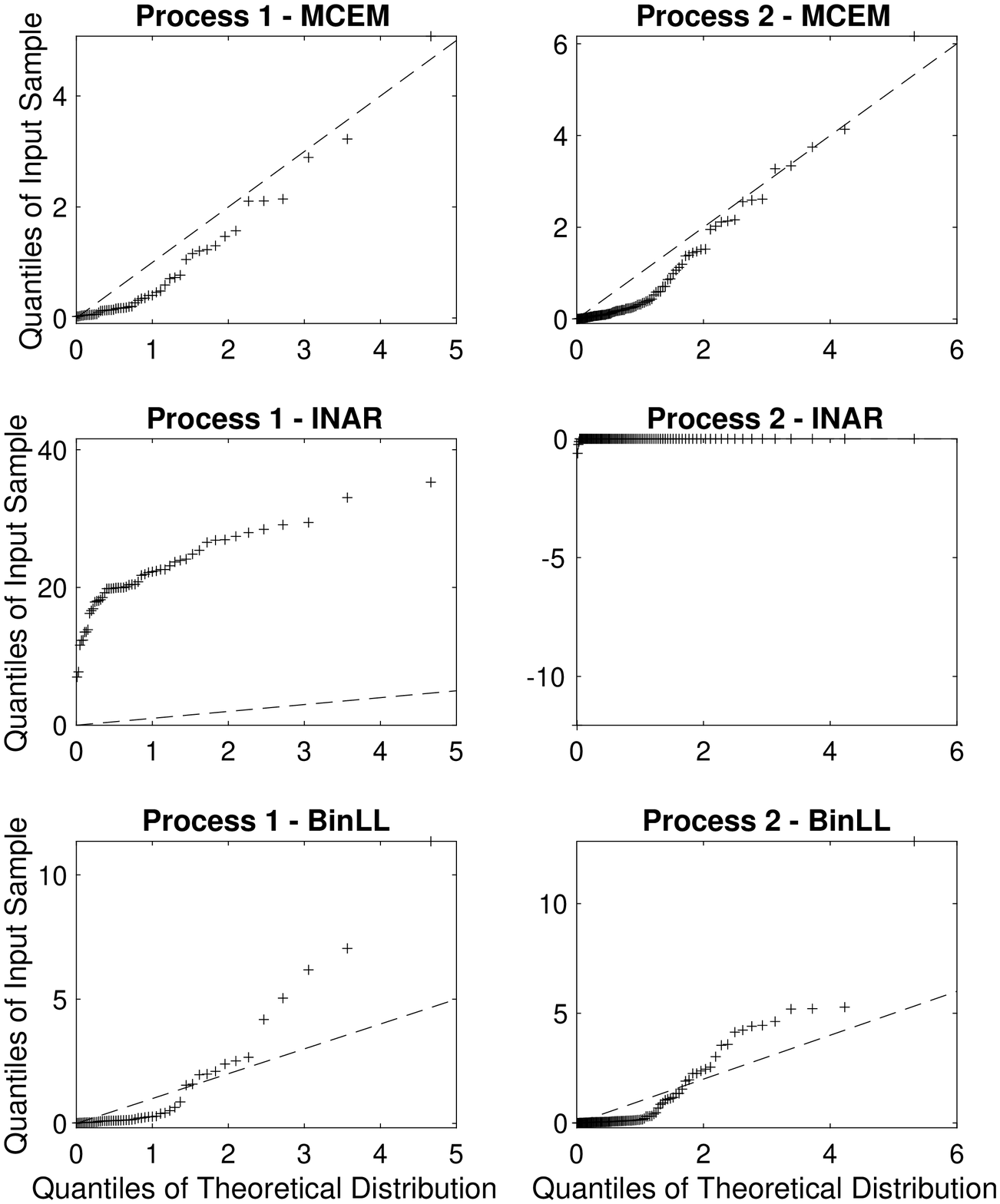}
	\caption{QQ-plots of transformed time-points using parameters estimated from each of the three methods for an edge in the LANL network. Note that the INAR($p$) method generates negative parameter estimates for process 2 and therefore has a particularly poor fit.} 
	\label{fig:11}
\end{figure*}

Parameter estimates are found using the MC-EM, INAR($p$) and binned log-likelihood methods and performance is considered using goodness of fit. This allows us to assess whether the estimates represent viable parameters for modelling the observed data as a mutually-exciting Hawkes process. To do this, we use the time rescaling theorem discussed in Section \ref{sec:sim_study}. 
Figure \ref{fig:11} shows that the parameter estimates obtained via the MC-EM algorithm are likely viable parameters. The MC-EM parameter estimates are 
\begin{equation*}
\bm{\nu}=
\left[ {\begin{array}{c}
	0.01 \\
	0.01 \\
	\end{array} } \right], \;  
\bm{\alpha}=
\left[ {\begin{array}{cc}
	0.49 & 0.00 \\
	0.22 & 0.28 \\
	\end{array} } \right], \; 
\bm{\beta}=
\left[ {\begin{array}{cc}
	1.46 & 0.53 \\
	0.80 & 1.01 \\
	\end{array} } \right].
\end{equation*}
The branching ratio, defined in Section \ref{sec:mv_hp_theory} is therefore
\begin{equation*}
\bm{\gamma}= \bm{\alpha} \oslash \bm{\beta} = 
\left[ {\begin{array}{cc}
	0.34 & 0.00 \\
	0.27 & 0.28 \\
	\end{array} } \right],
\end{equation*}
where $\oslash $ denotes element-wise division and $\gamma_{ij}$ is the average number of events in process $j$ directly triggered by each event in process $i$.

\begin{figure*}[!htb]
	\centering
	\includegraphics[scale=0.4]{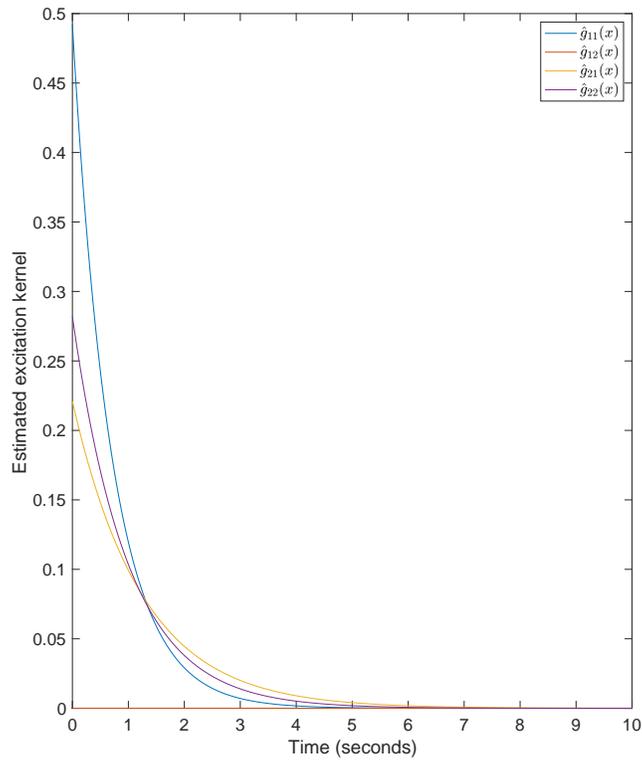}
	\caption{The excitation kernel ${\bm{g}}(u)$ as estimated by the MC-EM algorithm.} 
	\label{fig:12}
\end{figure*}
\FloatBarrier

These results indicate that there is mutually-exciting behavior between these processes in \textit{one direction} such that process 2 does not affect process 1 but process 1 does affect process 2, where process 1 is presented in blue in Figure \ref{fig:10} and process 2 in red. There parameter estimates suggest that for each event in process 1, an average of 0.27 events will be triggered in process 2. The baseline parameters given by $\bm{\nu}$ indicate the rate of the events is approximately one event every 100 seconds. Both processes are modelled with self-excitation, with each event in processes 1 and 2 triggering 0.34 and 0.28 events in their respective processes. Figure \ref{fig:12} shows the estimated excitation kernel components. From this we can see nature of the exciting effects, where $g_{ij}(x)$ illustrates the excitational effect which process $j$ has on process $i$.

\section{Conclusion}

Here we have presented a novel method for generalising an MCEM algorithm to handle multivariate aggregated data. By reparameterizing our multivariate model in terms of the superposed process we can inject the necessary cross-covariance structure required for generating valid proposals. We also present closed form expressions for the gradient and Hessian of the log likelihood for increased computational efficiency. We further conducted a simulation study to compare this approach to the INAR($p$) approximation detailed in \cite{kirchner_hawkes_2016} and a multivariate extension of the binned log likelihood method from \cite{mark_network_2019, shlomovich_monte_2020}. As in the univariate case, the MCEM method out-performed both alternatives in the presented parameter set and moreover, $\Delta$ can vary provided the interval bounds are known. The multivariate extension can also be applied for other Hawkes kernels.

\bibliographystyle{abbrvnat}
\bibliography{arxivMultivariate} 

\begin{thebibliography}{14}
\providecommand{\natexlab}[1]{#1}
\providecommand{\url}[1]{\texttt{#1}}
\expandafter\ifx\csname urlstyle\endcsname\relax
  \providecommand{\doi}[1]{doi: #1}\else
  \providecommand{\doi}{doi: \begingroup \urlstyle{rm}\Url}\fi

\bibitem[Bacry et~al.(2015)Bacry, Mastromatteo, and Muzy]{bacry_hawkes_2015}
E.~Bacry, I.~Mastromatteo, and J.-F. Muzy.
\newblock Hawkes processes in finance.
\newblock \emph{Market Microstructure and Liquidity}, 1\penalty0 (1), June
  2015.

\bibitem[Bowsher(2007)]{bowsher_modelling_2007}
C.~G. Bowsher.
\newblock Modelling security market events in continuous time: {Intensity}
  based, multivariate point process models.
\newblock \emph{Journal of Econometrics}, 141\penalty0 (2):\penalty0 876--912,
  Dec. 2007.

\bibitem[Daley and Vere-Jones(2003)]{daley_introduction_2003}
D.~J. Daley and D.~Vere-Jones.
\newblock \emph{An introduction to the theory of point processes}.
\newblock Springer, New York, 2nd edition, 2003.

\bibitem[Dempster et~al.(1977)Dempster, Laird, and
  Rubin]{dempster_maximum_1977}
A.~P. Dempster, N.~M. Laird, and D.~B. Rubin.
\newblock Maximum likelihood from incomplete data via the {EM} algorithm.
\newblock \emph{Journal of the Royal Statistical Society. Series B
  (Methodological)}, 39:\penalty0 1--38, 1977.

\bibitem[Hawkes(1971{\natexlab{a}})]{hawkes_point_1971b}
A.~G. Hawkes.
\newblock {\GG{2}}{P}oint spectra of some mutually exciting point processes.
\newblock \emph{Journal of the Royal Statistical Society. Series B
  (Methodological)}, 33\penalty0 (3):\penalty0 438--443, 1971{\natexlab{a}}.

\bibitem[Hawkes(1971{\natexlab{b}})]{hawkes_spectra_1971a}
A.~G. Hawkes.
\newblock {\GG{1}}{S}pectra of some self-exciting and mutually exciting point
  processes.
\newblock \emph{Biometrika}, 58\penalty0 (1):\penalty0 83--90,
  1971{\natexlab{b}}.

\bibitem[Kirchner(2016)]{kirchner_hawkes_2016}
M.~Kirchner.
\newblock Hawkes and {INAR}($\infty$) processes.
\newblock \emph{Stochastic Processes and their Applications}, 126\penalty0
  (8):\penalty0 2494--2525, Aug. 2016.

\bibitem[Kobayashi and Lambiotte(2016)]{kobayashi_tideh:_2016}
R.~Kobayashi and R.~Lambiotte.
\newblock {TiDeH}: {Time}-dependent {Hawkes} process for predicting retweet
  dynamics.
\newblock In \emph{Tenth {International} {AAAI} {Conference} on {Web} and
  {Social} {Media}}, Mar. 2016.

\bibitem[Mark et~al.(2019)Mark, Raskutti, and Willett]{mark_network_2019}
B.~Mark, G.~Raskutti, and R.~Willett.
\newblock Network estimation from point process data.
\newblock \emph{IEEE Transactions on Information Theory}, 65\penalty0
  (5):\penalty0 2953--2975, May 2019.

\bibitem[Ozaki(1979)]{ozaki_maximum_1979}
T.~Ozaki.
\newblock Maximum likelihood estimation of {Hawkes}' self-exciting point
  processes.
\newblock \emph{Annals of the Institute of Statistical Mathematics},
  31\penalty0 (1):\penalty0 145--155, Dec. 1979.

\bibitem[Price-Williams and Heard(2019)]{price-williams_nonparametric_2019}
M.~Price-Williams and N.~A. Heard.
\newblock Nonparametric self-exciting models for computer network traffic.
\newblock \emph{Statistics and Computing}, pages 1--12, May 2019.

\bibitem[Shlomovich et~al.(2020)Shlomovich, Cohen, Adams, and
  Patel]{shlomovich_monte_2020}
L.~Shlomovich, E.~Cohen, N.~Adams, and L.~Patel.
\newblock A {Monte} {Carlo} {EM} algorithm for the parameter estimation of
  aggregated {Hawkes} processes.
\newblock \emph{arXiv:2001.07160 [stat]}, Jan. 2020.
\newblock arXiv: 2001.07160.

\bibitem[Turcotte et~al.(2018)Turcotte, Kent, and Hash]{turcotte_unified_2018}
M.~J.~M. Turcotte, A.~D. Kent, and C.~Hash.
\newblock Unified host and network data set.
\newblock In \emph{Data {Science} for {Cyber}-{Security}}, Security {Science}
  and {Technology}, pages 1--22. World Scientific, Nov. 2018.
\newblock ISBN 978-1-78634-563-9.

\bibitem[Wei and Tanner(1990)]{wei_monte_1990}
G.~C.~G. Wei and M.~A. Tanner.
\newblock A {Monte} {Carlo} implementation of the {EM} algorithm and the {Poor}
  {Man}'s {Data} {Augmentation} algorithms.
\newblock \emph{Journal of the American Statistical Association}, 85\penalty0
  (411):\penalty0 699--704, 1990.

\end{thebibliography}

\appendix
\section{Gradient and Hessian of Multivariate Continuous Time Hawkes Processes}\label{app:likelihood}

The likelihood function of the multivariate Hawkes process is given by
\[
\log \mathcal{L}\left(\bm{\Theta} ; \bm{\mathcal{T}}\right)=\sum_{m=1}^{P} \log \mathcal{L}^{m}\left(\bm{\Theta} ; \bm{\mathcal{T}}\right)
\]
where
\[
\begin{aligned}
\log \mathcal{L}^{m}\left(\bm{\Theta} ; \bm{\mathcal{T}}\right)=-\nu_{m} T &-\sum_{n=1}^{P} \frac{\alpha_{m n}}{\beta_{m n}} \sum_{k: t_{k}^{n}<T}\left[1-\exp \left(-\beta_{m n}\left(T-t_{k}^{n}\right)\right)\right] \\
&+\sum_{k: t^m_{k}<T} \log \left(\nu_{m}+\sum_{n=1}^{P} \alpha_{m n} R_{m n}(k)\right)
\end{aligned},
\]
where $R_{m n}(k)$ is defined as
\[
R_{m n}(k)=\sum_{i: t_{i}^{n}<t_{k}^{m}} \exp \left(-\beta_{m n}\left(t_{k}^{m}-t_{i}^{n}\right)\right), \quad k \geq 2,
\]
and $R_{m n}(1) = 0$. This can recursively be defined as
\[
R_{m n}(k)=\exp ^{-\beta_{m n}\left(t_{k}^{m}-t_{k-1}^{m}\right)} R_{m n}(k-1)+\sum_{i: t_{k-1}^{m}<t_{i}^{n}<t_{k}^{m}} \exp \left(-\beta_{m n}\left(t_{k}^{m}-t_{i}^{n}\right)\right).
\]
Therefore, the gradient can be expressed by the following.
$$\frac{\partial L^{m}}{\partial \nu_{m}}=-T+\sum_{k: t_{k}^{m}<T} \frac{1}{\nu_{m}+\sum_{j=1}^{P} \alpha_{m j} R_{m j}(k)},$$

$$\frac{\partial L^{m}}{\partial \alpha_{m n}}=-\frac{1}{\beta_{m n}} \sum_{k: t_{k}^{n}<T}\left[1-\exp \left(-\beta_{m n}\left(T-t_{k}^{n}\right)\right)\right]+\sum_{k: t_{k}^{m}<T} \frac{R_{m n}(k)}{\nu_{m}+\sum_{j=1}^{P} \alpha_{m j} R_{m j}(k)}.$$

\begin{align*}
\frac{\partial L^{m}}{\partial \beta_{m n}}=&\frac{\alpha_{mn}}{\beta_{m n}^{2}} \sum_{k: t_{k}^{n}<T}\left[1-\exp \left(-\beta_{m n}\left(T-t_{k}^{n}\right)\right)\right]-\frac{\alpha_{m n}}{\beta_{m n}} \sum_{k: t_{k}^{n}<T}\left[\left(T-t_{k}^{n}\right) \exp \left(-\beta_{m n}\left(T-t_{k}^{n}\right)\right)\right]\\
&-\sum_{k: t_{k}^{m}<T} \frac{\alpha_{m n} R_{m n}^{\prime}(k)}{\nu_{m}+\sum_{j=1}^{P} \alpha_{m j} R_{m j}(k)},
\end{align*}
where
$$
R_{m n}^{\prime}(k)=\sum_{i: t^n_{i}<t^{m}_{k}}\left(t_{k}^{m}-t_{i}^{n}\right) \exp \left(-\beta_{m n}\left(t_{k}^{m}-t_{i}^{n}\right)\right), \quad k \geq 2,
$$
and $R_{m n}^{\prime}(1) = 0$. \\

We can also compute the Hessian for the continuous multivariate Hawkes likelihood. There are 15 categories to consider, given here. 
\begin{align*}
\frac{\partial^2 \mathcal{L}^m}{\partial \nu_m^2} = - \sum_{k: t_k^m < T} \frac{1}{\left(\nu_{m}+\sum_{j=1}^{P} \alpha_{m j} R_{m j}(k)\right)^{2}},
\end{align*}
\begin{align*}
\frac{\partial^2 \mathcal{L}^m}{\partial \nu_m \partial \nu_n} = 0, \quad m \neq n, 
\end{align*}
\begin{align*}
\frac{\partial^2 \mathcal{L}^m}{\partial \alpha_{mn}^2} =-\sum_{k: t_{k}^{m}<T}\left[\frac{R_{mn}(k)}{\nu_{m}+\sum_{j=1}^{P} \alpha_{m j} R_{m j}(k)}\right]^{2},
\end{align*}
\begin{align*}
\frac{\partial^2 \mathcal{L}^m}{\partial \alpha_{mn} \partial \alpha_{mn'}} = -\sum_{k: t_{k}^{m}<T} \frac{R_{m n}(k) R_{m n'}(k)}{\left(\nu_m+\sum_{j=1}^{P} \alpha_{m j} R_{mj}(k)\right)^{2}},
\quad n' \neq n,
\end{align*}
\begin{align*}
\frac{\partial^2 \mathcal{L}^m}{\partial \alpha_{mn} \partial \alpha_{m'n'}} = 0, \quad m' \neq m, \quad n,n' \in \{1, \ldots, p\},
\end{align*}
\begin{align*}
\frac{\partial^2 \mathcal{L}^m}{\partial \alpha_{mn} \partial \nu_{m}} = -\sum_{k: t_{k}^{m}<T} \frac{R_{mn}(k)}{\left( \nu_{m}+\sum_{j=1}^{P} \alpha_{m j} R_{m j}(k) \right)^2},
\end{align*}
\begin{align*}
\frac{\partial^2 \mathcal{L}^m}{\partial \alpha_{mn} \partial \nu_{m'}} = 0, \quad m' \neq m,
\end{align*}
\begin{align*}
\frac{\partial^2 \mathcal{L}^m}{\partial \beta_{mn} \partial \nu_{m}} = \sum_{k: t_{k}^{m}<T} \frac{\alpha_{m n} R_{m n}^{\prime}(k)}{\left(\nu_{m}+\sum_{j=1}^{P} \alpha_{mj} R_{m j}(k)\right)^{2}},
\end{align*}
\begin{align*}
\frac{\partial^2 \mathcal{L}^m}{\partial \beta_{mn} \partial \nu_{m'}} = 0, \quad m' \neq m, 
\end{align*}
\begin{align*}
\frac{\partial^2 \mathcal{L}^m}{\partial \beta_{mn} \partial \alpha_{mn}} &= 
-\frac{1}{\beta_{m n}} \sum_{k:t_{k}^{n}<T}\left[ \left(T-t_{k}^{n}\right) \exp \left(-\beta_{m n} \left(T-t_k^n\right)\right)\right] + \\
&\frac{1}{\beta_{m n}^{2}} \sum_{k: t_{k}^{n}<T}^{T}\left[1-\exp \left(-\beta_{m n}\left(T-t_{k}^{n}\right)\right)\right] - 
\sum_{k: t^m_{k}<T} \frac{R_{m n}^{\prime}(k)}{\nu_{m}+\sum_{j=1}^{P} \alpha_{m j} R_{m j}(k)} + \\
&\sum_{k: t^m_{k}<T} \frac{\alpha_{m n} R_{mn}^{\prime}(k) R_{m n}(k)}{\left(\nu_{m}+\sum_{j=1}^{P} \alpha_{m j} R_{m j}(k)\right)^{2}},
\end{align*}
\begin{align*}
\frac{\partial^2 \mathcal{L}^m}{\partial \beta_{mn} \partial \alpha_{mn'}} = \sum_{k: t_{k}^{m}<T} \frac{\alpha_{m n} R_{m n}^{\prime}(k) R_{m n'}(k)}{\left(\nu_{m}+\sum_{j=1}^{P} \alpha_{m j} R_{m j}(k)\right)^{2}}
\end{align*}
\begin{align*}
\frac{\partial^2 \mathcal{L}^m}{\partial \beta_{mn} \partial \alpha_{m'n'}} = 0, \quad m' \neq m, \quad n,n' \in \{1, \ldots, p\}, 
\end{align*}
\begin{align*}
\frac{\partial^2 \mathcal{L}^m}{\partial \beta^2_{mn}} = &-\frac{2 \alpha_{mn}}{\beta_{mn}^3} \sum_{k: t_{k}^{n}<T} \left[1-\exp \left(-\beta_{mn}\left(T-t_{k}^{n}\right)\right)\right] +\frac{2 \alpha_{mn}}{\beta_{mn}^{2}} \sum_{k: t_k^n<T} \left[\left(T-t_{k}^{n}\right) \exp \left(-\beta_{mn}\left(T-t_{k}^{n}\right)\right)\right]\\
&+\frac{\alpha_{m n}}{\beta_{m n}} \sum_{k: t_{k}^n<T}\left[\left(T-t_{k}^{n}\right)^{2} \exp \left(-\beta_{mn}\left(T-t_{k}^{n}\right)\right)\right]
+\sum_{k: t^m_{k}<T}\Biggl[\frac{\alpha_{m n} R_{mn}^{\prime \prime}(k)}{\nu_{m}+\sum_{j=1}^{P} \alpha_{m j} R_{m j}(k)}\\
&-\left(\frac{\alpha_{m n} R_{m n}^{\prime}(k)}{\nu_{m}+\sum_{j=1}^{P} \alpha_{m j} R_{m j}(k)}\right)^{2}\Biggr],
\end{align*}
where
\begin{align*}
R_{m n}^{\prime \prime}(k)=\sum_{i: t_{i}^{n}<t_{k}^{m}}\left(t_{k}^{m}-t_{i}^{n}\right)^{2} \exp \left(-\beta_{m n}\left(t_{k}^{m}-t_{i}^{n}\right)\right), \quad k \geq 2
\end{align*}
and $R_{mn}^{\prime \prime}(1) = 0.$
\begin{align*}
\frac{\partial^2 \mathcal{L}^m}{\partial \beta_{mn} \partial \beta_{mn'}} = -\sum_{k: t_{k}^{m}<T} \frac{\alpha_{m n} R_{m n}^{\prime}(k) \alpha_{m n'} R_{m n'}^{\prime}(k)}{\left( \nu_{m}+\sum_{j=1}^{P} \alpha_{mj}R_{mj}(k) \right)^2},
\quad n' \neq n,
\end{align*}
\begin{align*}
\frac{\partial^2 \mathcal{L}^m}{\partial \beta_{mn} \partial \beta_{m'n'}} = 0, \quad m' \neq m, \quad n,n' \in \{1, \ldots, p\}.
\end{align*}
This covers all cases required for the full Hessian matrix. 

\section{Multivariate MC-EM Algorithm}\label{app:mcem_alg}
Here we provide an algorithm for the parameter estimation of multivariate aggregated Hawkes processes via the MC-EM procedure.

\begin{algorithm}[ht]
	\caption{MC-EM}
	\label{pseudoMCEM}
	\begin{algorithmic}[1]
		\Function{MCEM}{${\bm{N}},M,\tilde{m},\epsilon$}
		\State $\tilde{N} \gets \sum_{p = 1}^P N^{(p)} $, to generate the superposed process, where $P$ is the dimension of $\bm{N}$ 
		\State $[\bm{\nu}^1, \bm{\alpha}^1, \bm{\beta}^1] = \Theta^1 \gets \rm{Unif}(P,1 + 2P)$ such that the spectral radius, $\rho(\bm{\gamma})<1$ to ensure stationarity
		\State $i \gets 1$
		\While{tolerance $> \epsilon$}
		\State $\tilde{\Theta}^i = [\tilde{\bm{\nu}}^i, \tilde{\bm{\alpha}}^i, \tilde{\bm{\beta}}^i]$, the corresponding superposed estimate, is formed using the reparameterization given in Section \ref{sec:superpose}
		\For{$j = 1$ to $M$}
		\State $\tilde{\mathcal{T}}^{\ast(j)} \sim q(\tilde{\mathcal{T}} \mid {\tilde{\bm{N}}}, \tilde{\Theta}^i)$, generate univariate proposal times as in Algorithm 1 from \cite{shlomovich_monte_2020}
		\For{$l = 1$ to $\tilde{m}$}
		\State Uniformly sample without replacement the observed number of points for each bin, for each process from $\tilde{\mathcal{T}}^{\ast(j)}$ to form $\bm{\mathcal{T}}^{\ast(j,l)}$. This is the MC sample of the multivariate latent times $\bm{\mathcal{T}}$, using the $j$th MC sample of the superposed latent times $\tilde{\mathcal{T}}$
		\EndFor 
		\State $\bm{\mathcal{T}}^{\ast (j)} \gets \bm{\mathcal{T}}^{\ast(j,l)}$ such that $l = \argmax_{l} \left( \log \left( p \left( \bm{\mathcal{T}}^{\ast (j,l)} \mid \Theta^i \right) \right) \right)$ 
		\State $w_j \gets \log \left( p(\bm{\mathcal{T}}^{\ast (j)} \mid \Theta^i)/q(\bm{\mathcal{T}}^{\ast (j)} \mid {\bm{N}}, \Theta^i) \right) $
		\EndFor
		\State ${\bm{w}} = {\bm{w}} - C$, scale the weights with an appropriately chosen $C$, such as $C \approx \min(\bm{w})$
		\State $Q_{i+1}(\bm{\Theta}, \bm{\Theta}^{i}) \gets \sum_{k=1}^M w_k \log (p (\bm{\Theta} \mid {\bm{N}}, \bm{\mathcal{T}}^{\ast})) / \sum_{k=1}^M w_k$
		\State $\Theta^{i+1} \gets \argmax_{\bm{\Theta}, \rho(\gamma)<1} Q_{i+1}(\bm{\Theta}, \bm{\Theta}^{i}) $	 
		\State tolerance $\gets {\rm{norm}}(\bm{\Theta}^{i+1} - \bm{\Theta}^{i}) $
		\State $i \gets i + 1$
		\EndWhile
		\State \textbf{return} $\{\bm{\Theta}^i\}$ \Comment{Set of parameter estimates}
		\EndFunction 
	\end{algorithmic}
\end{algorithm}

\end{document}